\documentclass[11pt]{amsart}
\usepackage{amsbsy,amssymb,amscd,amsfonts,latexsym,amstext,delarray, amsmath,graphicx,color,caption}
\usepackage{tikz}
\usetikzlibrary{matrix,arrows}
\usepackage{graphicx}
\input xy

\xyoption{all}
\usepackage{euscript}
\usepackage{multirow}
\usepackage{etex, pictexwd,dcpic}

\def\to{\longrightarrow}

\newcommand{\Z}{\mathbb{Z}}

\newtheorem{theorem}{Theorem}[section]

\def\Bbb#1{\mathbb{#1}}

\def\C{{\mathbb C}}

\def\R{{\mathbb R}}
\def\Z{{\mathbb Z}}

\def\a{\alpha}

\def\Mg{\mathfrak{g}}

\begin{document}

\title{\bf {Kac--Moody groups and automorphic forms in low dimensional  supergravity theories}}

\date{\today}

\author{Ling Bao and  Lisa Carbone}

\vspace{5.0cm}

\begin{abstract} Kac--Moody groups $G$ over $\mathbb{R}$ have been conjectured to occur as symmetry groups of supergravity theories dimensionally reduced to  dimensions less than 3, and their integral forms $G(\mathbb{Z})$ conjecturally encode quantized symmetries.   In this review paper, we briefly introduce the conjectural symmetries of Kac--Moody groups in supergravity as well as the known evidence for these conjectures. We describe constructions of Kac--Moody groups over $\R$ and $\Z$ using certain choices of fundamental modules that are considered to have physical relevance. Eisenstein series on certain finite dimensional algebraic groups are known to encode quantum corrections in the low energy limit of superstring theories. We describe briefly how the construction of Eisenstein series extends to Kac--Moody groups.  The constant terms of Eisenstein series on  $E_9$, $E_{10}$ and $E_{11}$ are predicted to encode perturbative string theory corrections.

\end{abstract}

\maketitle

\section{Introduction}

 Kac--Moody groups and algebras are the most natural generalizations to infinite dimensions of finite dimensional simple Lie groups and Lie algebras. Affine Kac--Moody algebras and their generalizations by Borcherds have concrete physical realizations and have wide applications in physical theories. Suitable extensions of the Dynkin diagrams of affine Kac--Moody algebras give rise to hyperbolic and Lorentzian Kac--Moody algebras, such as $E_{10}$ and $E_{11}$. 
 
As discussed in some of the talks,  hyperbolic and Lorentzian Kac--Moody groups and algebras have recently been discovered as symmetries in  supergravity theories.  To understand the group theoretic questions that arise, it's instructive to first briefly view the symmetries in a physical context.  An excellent and detailed survey of the physics we will discuss is given in [FKGP]  (see also [D1], [DN], [HPS], [FKP], [KN], [Ni2] and [Ni3]).  An expanded version of this article is available at [BC].

We describe a construction of a representation theoretic  Kac--Moody group $G_V$,   over $\R$ and $\Z$ using integrable highest weight modules $V$ and a $\Z$--form of the universal enveloping algebra. We describe a choice of fundamental modules $V$ for the Kac--Moody groups $E_9$, $E_{10}$ and $E_{11}$ that are conjectured to play a role in certain supergravity theories.

  Let  $G$ be  (finite dimensional) semisimple algebraic group and let $K$ be a maximal compact  subgroup of $G$. Eisenstein series  on $K\backslash G(\R)/G(\Z)$ appear naturally in the context of superstring theories, where they encode  quantum corrections in the low energy regime described by  supergravity theories ([GMRV], [LW]). 
  
We describe briefly how the construction of Eisenstein series extends to Kac--Moody groups. It is also thought that the Eisenstein series on the Kac--Moody groups $E_9$, $E_{10}$ and $E_{11}$ should encode the quantum correction terms in string theory ([FK]). We outline the construction of Eisenstein series on these groups and we review current developments and open questions in the study of automorphic forms on non--affine Kac--Moody groups.

 \section{Supergravity theories}
 
 The leading candidate in theoretical physics for a single consistent theory of all  the fundamental forces is {\it superstring theory}. However, there is no single theory of superstrings, but rather five theories that describe various aspects of the behaviour of strings.  In the 1990's, Witten, building on previous insights by Duff, Townsend, and others,  proposed an overarching theory, known as {\it M--theory}, whose various special limits are the superstring theories  ([Wi1], [Wi2]). M--theory  has a  rich structure, both mathematically and physically, but has not yet been fully formulated and will likely require new mathematics  for its full development.  

 Essential for  superstring theory is {\it supersymmetry}. This  symmetry, discovered in 1972,  transforms fermions into bosons and vice versa.  {\it Supergravity} is a theory that incorporates  both quantum field theory and general relativity using supersymmetry. It was discovered in 1978 as a 4 dimensional theory, but it can occur in dimensions up to 11, where there are 7 additional spacial dimensions.

  The possible supergravity theories were classified  in the 1970s. Given a physical theory (Lagrangian) in $D$ dimensions, {\it dimensional reduction} is a procedure that gives rise to a Lagrangian in $d<D$ dimensions, by taking all fields to be independent of location in the extra $D-d$ dimensions.  The extra dimensions are contracted to an `invisible scale' along a choice of vector field, involving the full geometry of spacetime.  

 The type IIA and type IIB supergravity theories in
10--dimensions are the low energy limit of the
type II string theories while the low energy limit of M--theory is known to coincide with 11--dimensional supergravity. These supergravity theories are important, as they encode both perturbative and non--pertubative
effects, many of which are not easily calculated from first principles in string theory or M--theory (see for example [GLW]).

The {\it maximal supergravities} are those with the maximal number of {\it supercharges}, or fermionic generators of the  super--Poincar\'e algebra.  The maximal supergravities in dimensions less than 11 are obtained by dimensional reduction of the classical action of 11--dimensional supergravity on an $n$--torus. Dimensional reduction may be carried out on other compact manifolds, but on an $n$--torus  all the supercharges are preserved. The resulting $(11-n)$--dimensional supergravity action in the Einstein frame has the property that the scalar fields of the maximal supergravity theory in $(11-n)$ dimensions take values in the coset  $K(E_{n}(\R)) \backslash E_{n}(\R)$, where $E_n$  is a split real form of the simple exceptional Lie group of type $E_n$ and  $K(E_n)$ is a subgroup invariant under the Cartan involution. 

For example, on dimensional reduction to  $D=4$ spacetime dimensions, 11--dimensional supergravity has the  maximal  number $N=8$ of supercharges and the scalar fields take values in the coset $K(E_{7}(\R)) \backslash E_{7}(\R)$. We discuss this example in more detail in the next subsection.

Symmetries of the scalar cosets were first established for $1 \le n \le 8$ in [CJ]. For $n=9$,  it was discovered that the scalar fields take values in the  coset  $K(E_9) \backslash E_{9}(\R)$ arising from the affine Kac--Moody group $E_9$  ([Ju1] and [N]). Moreover, the supergravity equations of motion are invariant under the Virasoro algebra.

 A correspondence between the scalar fields of 11--dimensional supergravity theory dimensionally reduced to 1 dimension and the hyperbolic $E_{10}$ coset $K(E_{10}) \backslash E_{10}(\R)$ was established in [DHN1] after certain truncations were made on both sides of the correspondence, in particular,  only at `low levels' of the roots of $E_{10}$. The subgroup  $E_{10}(\mathbb{Z})$ of  $E_{10}(\mathbb{R})$ has been conjectured to be a discrete symmetry group of Type II superstring theory ([Ju1], [HT]).

 West  ([W1]) showed that truncated versions of the bosonic sectors of 11--dimensional supergravity and type IIA supergravity can be derived in terms of a certain truncation of the Lorentzian $E_{11}$ coset $K(E_{11}) \backslash E_{11}$.

 The full list of symmetry groups in maximal supergravity theories is summarized in Table \ref{tab:coset}.
\begin{table}
  \begin{center}
    \begin{tabular}{l|lll}
      Dimension & $K(G)$              					& $G(\R)$                     & $G(\Z)$ \\
      \hline
      10, IIA   & \textbf{1}          					& $\R^+$          				& \textbf{1} \\
      10, IIB   & $SO(2)$             					& $SL_2(\R)$                  & $SL_2(\Z)$ \\
      9         & $SO(2)$      		  					& $SL_2(\R)\times \R^+$ 		& $SL_2(\Z)$ \\
      8         & $SO(3)\times SO(2)$ 					& $SL_3(\R)\times SL_2(\R)$ 	& $SL_3(\Z)\times SL_2(\Z)$ \\
      7         & $SO(5)$             					& $SL_5(\R)$                  & $SL_5(\Z)$ \\
      6         & $(Spin(5)\times Spin(5))/\Z_2$ 	& $Spin(5,5;\mathbb{R})$      & $Spin(5,5;\mathbb{Z})$ \\
      5         & $USp(8)/\Z_2$       					& $E_{6}(\mathbb{R})$         & $E_{6}(\mathbb{Z})$ \\
      4         & $SU(8)/\Z_2$        					& $E_{7}(\mathbb{R})$         & $E_{7}(\mathbb{Z})$ \\
      3         & $Spin(16)/\Z_2$     					& $E_{8}(\mathbb{R})$         & $E_{8}(\mathbb{Z})$ \\
		2         & $K(E_9)$            					& $E_{9}(\mathbb{R})$         & $E_{9}(\mathbb{Z})$ \\
		1         & $K(E_{10})$         					& $E_{10}(\mathbb{R})$        & $E_{10}(\mathbb{Z})$ \\
		0         & $K(E_{11})$         					& $E_{11}(\mathbb{R})$        & $E_{11}(\mathbb{Z})$
    \end{tabular}
  \end{center}
  \caption{Coset symmetries in maximal supergravities.}
  \label{tab:coset}
\end{table}

\subsection{Example: maximal supergravity in $D=4$ spacetime dimensions}


It is known that the 11--dimensional maximal supergravity, dimensionally reduced to $D=4$ spacetime dimensions, has   global $E_{7}(\R)$ symmetry ([CJ]). That is, the equations of motion of the  supergravity theory, dimensionally reduced to $D=4$  are invariant  under $E_{7}(\R)$. However, the Lagrangian is not invariant under $E_{7}(\R)$,  but under a smaller group which acts on the vector fields.
The 70 scalar fields of this theory take values in the coset $[SU(8),\mathbb{R})/ \{\pm Id\}]\backslash E_{7}(\mathbb{R})$ ([CJ]).

In $D=4$ spacetime dimensions, the electric and magnetic charges of the theory lie  in a 56--dimensional vector space $V$, with 28 electric and 28 magnetic fundamental charges. These  charges are  subject to the `Dirac--Schwinger--Zwanziger quantization
condition', which constrains the charges to  lie on a  lattice  $V_{\Z}$ in  $V$ called the {\it charge lattice} ([HT]).

The space $V$ is the 56--dimensional fundamental representation of $\mathfrak{e}_7$. This representation gives rise to a faithful representation of $E_{7}(\R)$ in $Sp_{56}(\mathbb{R})$. The charge lattice admits an action of the discrete group $E_{7}(\mathbb{Z})$ preserving the set of electric and magnetic charges ([HT]).

Although the symmetry groups of the classical maximal supergravities involve Lie  groups constructed over the real numbers $\R$, adding quantum corrections is conjectured to break these continuous groups to discrete subgroups defined over $\Z$ ([HT]).  For 11--dimensional maximal supergravity, dimensionally reduced to $D=4$, the discrete group is $E_{7}(\mathbb{Z})=E_{7}(\mathbb{R}) \cap Sp_{56}(\mathbb{Z})$. Soul\'e gave a rigorous mathematical proof that the $E_{7}(\mathbb{Z})$ of [HT] coincides with the Chevalley $\Z$--form $G(\Z)$ of $G=E_7$ ([S]). 
In the appendix of [MS], the authors gave a physics inspired  construction of the group $E_8(\Z)$.

\section{Construction of Kac--Moody groups  over $\R$ and $\Z$}

  Now let $\Mg$ be a Kac--Moody algebra over $\C$ with simple roots $\alpha_i$, $i\in I$.  Here we summarize the construction of a Kac--Moody group $G$ associated to $\Mg$ in analogy with Chevalley's construction of finite dimensional semisimple algebraic groups. The generalization of this method to Kac--Moody groups over fields is provided by [CG] and over $\Z$ by [Ca].  Some external data is needed: namely an integrable highest weight module $V$ for $\Mg$ and a $\mathbb{Z}$-form $V_{\mathbb{Z}}$ constructed from a $\mathbb{Z}$-form of the universal enveloping algebra.

Let $V=V^{\lambda}$ be the unique irreducible highest weight module for $\mathfrak{g}$ corresponding to dominant integral weight $\lambda$. 
Let $\Lambda\subseteq \mathfrak{h}^*$ be the linear span of  the simple roots $\alpha_i$, for $i\in I$, and let $\Lambda^{\vee}\subseteq
\mathfrak{h}$ be the linear span of the simple coroots
$\alpha^{\vee}_i$, for $i\in I$.

Let $e_i$ and $f_i$ be the Chevalley generators of $\mathfrak{g}$. Let ${\mathcal U}_{\mathbb{C}}$ be the universal enveloping algebra of $\mathfrak{g}$. Let  ${\mathcal U}_{{\mathbb{Z}}}\subseteq {\mathcal U}_{{\mathbb{C}}}$ be the ${\mathbb{Z}}$--subalgebra generated by 
$$\dfrac{e_i^{m}}{m!},\ \dfrac{f_i^{m}}{m!},\ \left (\begin{matrix}
h\\ m\end{matrix}\right )$$ for $i\in I$, 
$h\in\Lambda^{\vee}$ and
$m\geq 0$.

Let  $v_{\lambda}\in V$ be a highest weight vector.  We set 
$$V^{\lambda}_{\mathbb{Z}}\  =\ \mathcal{U}_{\mathbb{Z}}\cdot v_{\lambda}.$$
Then $V^{\lambda}_{\mathbb{Z}}$  is a lattice in $V^{\lambda}_{R}=R\otimes_{\mathbb{Z}}V^{\lambda}_{\mathbb{Z}}$ and a ${\mathcal U}_\mathbb{Z}$--module.
 
 For $s,t\in R$ and $i\in I$,  set 
$$\chi_{\alpha_i}(t)\ =\ exp(\rho(se_i)),$$
$$\chi_{-\alpha_i}(t)\ =\ exp(\rho(tf_i)),$$
where $\rho$ is the defining representation for $V$.
Then these are elements of $Aut(V^{\lambda}_{R})$, thanks to the
local
nilpotence of $e_{i},$ $f_{i}.$

We let $G^{\lambda}(R)\leq Aut(V^{\lambda}_{R})$ be the group: 
$$G^{\lambda}(R)=\langle \chi_{\alpha_i}(s),\ \chi_{-\alpha_i}(t)\mid i\in I,\ s,t\in R\rangle.$$ 

\newpage
We summarize the construction as follows.

\begin{theorem} Let $\mathfrak{g}$ be a symmetrizable Kac--Moody algebra over a commutative ring $R$ with 1. Let  $\alpha_i$, $i\in I$, be the simple roots and $e_i$, $f_i$ the Chevalley generators of $\mathfrak{g}$. Let $V^{\lambda}_R$ be an $R$--form of an integrable highest weight module $V^{\lambda}$ for $\mathfrak{g}$, corresponding to  dominant integral weight $\lambda$ and defining representation $\rho:\mathfrak{g}\to End(V^{\lambda}_R)$. Then 
$$G^{\lambda}(R)=\langle  \chi_{\a_i}(s)=exp(\rho(se_i)),\ \chi_{-\a_i}(t)=exp(\rho(tf_i))\mid s,t\in R\rangle\leq Aut(V^{\lambda}_R)$$
is a representation--theoretic  Kac--Moody group associated to $\mathfrak{g}$.
\end{theorem}

Now we define Kac--Moody groups over $\R$ and $\Z$ as follows:
$$G^{\lambda}(\R)=\langle\chi_{\a_i}(s),\chi_{-\a_i}(t)\mid s,t\in\R,\ i\in I\rangle.$$
$$G^{\lambda}(\Z)=\langle\chi_{\a_i}(s),\chi_{-\a_i}(t)\mid s,t\in\Z,\ i\in I\rangle.$$
The following was  proven in [Ca]. 

\begin{theorem} ([Ca]) The group $G^{\lambda}(\Z)$ has the following  generating sets:

(1) $\chi_{\alpha_i}(1)$ and $\chi_{-\alpha_i}(1)$, 

and

(2) $\chi_{\alpha_i}(1)$ and $\widetilde{w}_{\alpha_i}(1)=\chi_{\alpha_i}(1)\chi_{-\alpha_i}(-1)\chi_{\alpha_i}(1)$.
\end{theorem}

The generating set (2) is the analog of the $S,T$-generating set for $SL_2(\Z)$ and coincides with the generating set obtained in [AC] where the authors also gave a finite presentation for the group $E_{10}(\Z)$.

\subsection{Explicit construction of the groups $E_9$, $E_{10}$ and $E_{11}$} 

An important question in string theory is to determine which representations of the underlying Kac--Moody algebras (or Lie algebras) enter into the construction of the automorphic forms. Here we describe possible choices of modules $V$ for $E_9$, $E_{10}$ and $E_{11}$ that have physical relevance.

 We summarize the constructions of the groups $E_{n}(\mathbb{R})$ and $E_{n}(\mathbb{Z})$ for $n=9,10,11$. In each of these constructions, we use the data in Table~\ref{tab:data}:

\begin{table}[ht!]
\caption{}
\centering
\begin{tabular}{| c | c |  c  | c | c|}
\hline
Algebra & Generators  & Simple roots & Fund. weights & Highest weight module\\
\hline

   $\mathfrak{e}_{9}(\mathbb{C})$
& 
$\begin{matrix}
& e_1,\dots ,e_{9}\\
& f_1,\dots ,f_{9}
\end{matrix}$

&
$
 \alpha_1,\dots ,\alpha_{9}
$

&
 $\omega_1,\dots ,\omega_{9}$
 
&  $\begin{matrix}
\text{$V=V^{\omega_{1}}$  }\\ 
\text{$V$ integrable  with high. wt. vec. $v^{\omega_1}$ }\\
\text{corresp. to fund. weight $\omega_{1}$ }
 \text{labeled as in}\\
  \text{Figure~\ref{fig:DynkinE11}}
\end{matrix}$

\\ \hline

   $\mathfrak{e}_{10}(\mathbb{C})$
& 
$\begin{matrix}
& e_1,\dots ,e_{10}\\
& f_1,\dots ,f_{10}
\end{matrix}$

&
$
 \alpha_1,\dots ,\alpha_{10}$

&
 $\omega_1,\dots ,\omega_{10}$
 
&  $\begin{matrix}
\text{$V=V^{\omega_1+\dots +\omega_{10}}$  }\\ 
\text{$V$ irred. and integrable with high. wt. vec. }\\
v^{\omega_1+\dots +\omega_{10}} \\
\end{matrix}$

\\ \hline

   $\mathfrak{e}_{11}(\mathbb{C})$
& 
$\begin{matrix}
& e_1,\dots ,e_{11}\\
& f_1,\dots ,f_{11}
\end{matrix}$

&
$\begin{matrix}
 \alpha_1,\dots ,\alpha_{11}\\
 \text{labeled as in}\\
  \text{Figure~\ref{fig:DynkinE11}}
\end{matrix}$

&
 $\omega_1,\dots ,\omega_{11}$
 
&  $\begin{matrix}
\text{$V=\omega_{11}$--rep. of $\mathfrak{e}_{11}(\mathbb{C})$.  }\\ 
\text{$V$ integrable  with high. wt. vec. $v^{\omega_{11}}$ }\\
\text{corresp. to fund. weight $\omega_{11}$ }
\end{matrix}$

\\ \hline
\end{tabular}
\label{tab:data}
\end{table}

With this data, we have the groups:
 
 $$\boxed{E_{n}(\mathbb{R})=\langle  \chi_{\alpha_i}(s)=exp(se_i)$, $\chi_{-\alpha_i}(t)=exp(tf_i) \mid s,t\in\mathbb{R}\ i=1,\dots ,n\rangle}$$

 $$\boxed{E_{n}(\mathbb{Z})=E_{n}(\mathbb{R})\cap Aut(V_{\mathbb{Z}})=\langle  \chi_{\alpha_i}(s)=exp(se_i)$, $\chi_{-\alpha_i}(t)=exp(tf_i) \mid s,t\in\mathbb{Z},\ i=1,\dots ,n\rangle}$$

\newpage

We now outline the choice of representation for the above group constructions.

\medskip
{\it {$\circ$ $E_{9}(\mathbb{Z})$ and $E_{9}(\mathbb{R})$} }

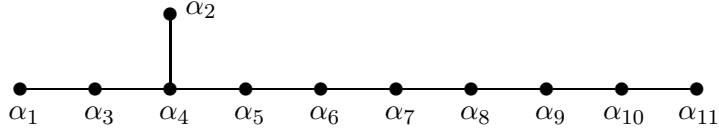
\begin{figure}
\begin{center}
\setlength{\unitlength}{1mm}
\begin{picture}(90,10)
   \put(0,0){\circle*{1.7}}
   \put(10,0){\circle*{1.7}}
   \put(20,0){\circle*{1.7}}
   \put(30,0){\circle*{1.7}}
   \put(40,0){\circle*{1.7}}
   \put(50,0){\circle*{1.7}}
   \put(60,0){\circle*{1.7}}
   \put(70,0){\circle*{1.7}}
   \put(80,0){\circle*{1.7}}
   \put(90,0){\circle*{1.7}}
   \put(20,10){\circle*{1.7}}

   \put(0,0){\line(1,0){90}}
   \put(20,0){\line(0,1){10}}

   \put(-1.5,-4){$\alpha_1$}
   \put(8.5,-4){$\alpha_3$}
   \put(22,10){$\alpha_2$}
   \put(18.5,-4){$\alpha_4$}
   \put(28.5,-4){$\alpha_5$}
   \put(38.5,-4){$\alpha_6$}
   \put(48.5,-4){$\alpha_7$}
   \put(58.5,-4){$\alpha_8$}
   \put(68.5,-4){$\alpha_9$}
   \put(77.5,-4){$\alpha_{10}$}
   \put(87.5,-4){$\alpha_{11}$}
 
\end{picture}

\end{center}
\bigskip
  \caption{The Dynkin diagram of $E_{11}$}
  \label{fig:DynkinE11}
\end{figure}

For $E_9$, we choose $V$ to be the fundamental representation $V^{\omega_1}$ which is the highest weight module with highest weight $\omega_1$, where $\omega_1$ is the fundamental weight dual to $\alpha_1$ as labeled in Figure~\ref{fig:DynkinE11}, viewing the Dynkin diagram for $E_9$ as a subdiagram of $E_{11}$. Our motivation for this is the following. Certain coefficients of quantum correction terms (the $\mathcal{R}^4$ and $\partial^4\mathcal{R}^4$ correction terms for instance) in type IIB supergravity, dimensionally reduced to $D\geq 3$, can be expressed in terms of Eisenstein series on various finite dimensional Lie groups. In particular, the coefficient of the $\mathcal{R}^4$ correction term is an Eisenstein series whose complex parameter is  a scalar multiple of $\omega_1$ ([GMRV]). It is an open question to determine if this symmetry structure extends to 2 dimensions with corresponding Eisenstein series on $E_9$.

\bigskip

{\it {$\circ$ $E_{10}(\mathbb{Z})$ and $E_{10}(\mathbb{R})$}}

When the root lattice $Q$ equals the weight lattice $P$, as is the case for $\mathfrak{e}_{10}$, that is, $|det(A)|=1$, where $A$ is the generalized Cartan matrix, then we may choose 
$V=V^{\omega_1+\dots +\omega_{10}}$, where $\omega_i$ are the fundamental weights. Then $V$ is an irreducible integrable highest weight module with lattice of weights equal to $P$ and with highest weight $\omega_1+\dots +\omega_{10}$. We have 
$$wts(V)\subseteq \{\omega_1+\dots +\omega_{10}-\sum_{j=1}^{10} k_{j}\alpha_j\mid k_{j}\in\mathbb{Z}_{\geq 0}\},$$
where $\alpha_i$ are the simple roots, and
$$\left(\begin{matrix}
\omega_1 \\
\omega_2\\
\vdots\\
\omega_{10}
\end{matrix}\right)=A^{-1}
\left(\begin{matrix}
\alpha_1 \\
\alpha_2\\
\vdots\\
\alpha_{10}
\end{matrix}\right).$$
Since the root lattice $Q=\mathbb{Z}\alpha_1\oplus \dots \oplus \mathbb{Z}\alpha_{10}$ and the weight lattice $P$ coincide, the weights of $V$ are contained in the $\mathbb{Z}$-span of the simple roots. Hence $wts(V)$ contains all the fundamental weights $\omega_i$. 

\newpage

\bigskip
{\it{$\circ$ $E_{11}(\mathbb{Z})$ and $E_{11}(\mathbb{R})$}}

 Symmetries of the discrete group $E_{11}(\mathbb{Z})$ were discussed in [GW1], where the authors conjecture that the group $E_{11}(\mathbb{Z})$ preserves the brane charge lattice of Type II superstring theory. This charge lattice belongs to the fundamental representation $V$ corresponding to the vertex at the end of the long tail of the Dynkin diagram for $E_{11}$. We call this representation  the $\omega_{11}$--representation. In [GW1], due to a different labeling of the vertices of the Dynkin diagram, this is called the $\ell_1$--representation.

We note that the above constructions depend on  choices of modules for which there is a physical motivation. These choices give rise to group constructions  which are useful in the physical contexts described above (for example our generating set for $E_{11}$ was used in [GW1]). 

In general, the dependence of the representation theoretic construction of Kac--Moody groups on the choice of module is not completely understood. Garland gave a representation theoretic construction of  affine Kac--Moody groups as central extensions of loop groups, where each central extension corresponds to a unique cohomology class represented by a cocycle, known as the {\it Steinberg cocycle} restricted to the torus $H$. Garland characterized  the dependence on the choice of highest weight module $V$ for affine groups in terms of the Steinberg cocycle.
In [CW], the authors proved that for different choices of highest weight module $V$, the discrepancy between the groups $E_{10}^V(\Z)$ is contained in a  finite abelian group of order at most $(\Z/2\Z)^{10}$.

\section{Eisenstein series in supergravity}

The first quantum corrections to the classical supergravity action were found by computing the scattering amplitudes of four closed interacting strings. 
In particular, these scattering amplitudes give rise to quantum corrections of the form $\mathcal{R}^4$ and $\partial^4\mathcal{R}^4$. The coefficient of the $\mathcal{R}^4$ correction term is an Eisenstein series whose complex parameter is  a scalar multiple of $\omega_1$ ([GMRV]).

Schematically, this scattering amplitude is separable into analytic and non--analytic parts
 $$
A = A^{\text{analytic}} + A^{\text{non-analytic}}.
$$
The analytic part of the scattering amplitude in $d=(11-n)$ dimensions is an expansion in automorphic forms on the double coset $K(E_n(\R)) \backslash E_{n}(\R)/E_{n}(\Z)$ ([GMRV]). Each automorphic form has a Fourier expansion, which naturally separates the perturbative and non--perturbative quantum corrections. 

Being the perturbative limit, the scattering amplitudes only capture the constant terms (zero--th order terms in the Fourier expansion) of the automorphic forms ([GRV]).

 The remaining parts of the Fourier expansion correspond to `non--perturbative instanton corrections' ([GG], [GV], [GMV]). The direct determination of instanton effects in string theory is usually very difficult. The benefit here is that the action of the integral group $G(\Z)$ mixes perturbative and
non--perturbative effects, which allows indirect computation of the instanton corrections. It is conjectured  that different types of instanton effects correspond to Fourier expansions of Eisenstein series with respect to different parabolic subgroups ([GMV], [FK]). 

By dimensionally reducing the higher derivative corrections of ten--dimensional IIB
theory on a torus, the authors of [GLW] and [GW1] describe some natural constraints on the $E_{n+1}$
automorphic forms that occur in $d = 10-n$ dimensions. They also argue that these automorphic forms involve the
representation of $E_{n+1}$ with highest weight $\omega_1$.

With regard to Eisenstein series on Kac--Moody groups, in analogy with $SL_2(\mathbb{Z})$, the group $E_{10}(\mathbb{Z})$ is conjectured to be a `modular group' for certain automorphic forms that are expected to arise in the context of 11 dimensional supergravity ([DN], [Ga]). Automorphic forms on $E_{10}$ and $E_{11}$  are conjectured  to encode  higher derivative corrections of string theory and M--theory ([DN2], [DHHKN], [W1]). The constant terms of Eisenstein series on $E_9$, $E_{10}$ and $E_{11}$ have been studied in [Fl]  in relation to string scattering amplitudes. We discuss this further in the next section.

\section{Eisenstein series on non--affine Kac--Moody groups}\label{KMEseries}

In his seminal work \cite{G99, G04, G06, G11, GMS1, GMS2, GMS3, GMS4}, Garland extended the classical theory of Eisenstein series to arithmetic  quotients  $K(G(\R))\backslash G(\R)/G(\Z) $ of affine Kac--Moody groups $G$. He proved absolute convergence in a half space and meromorphic continuation to the full complex plane. Garland's theory of Eisenstein series has been extended to affine groups over fields such as number fields ([Li]) and function fields ([BK], [LL], [P]) and to hyperbolic and more general Kac--Moody groups ([CLL], [CGLLM]).

 We outline the construction of general Eisenstein series on Kac--Moody groups. Let $G=G(\R)$ be a non--affine Kac--Moody group associated to a Kac--Moody algebra $\mathfrak{g}$. Let  $G(\R)=KA^+N$ be the Iwasawa decomposition of $G$, where  $K$ is the fixed point subgroup of the involution on  $G(\mathbb{R})$ induced from the Cartan involution on $\mathfrak{g}$, $A^+\cong (\mathbb{R}^+)^{rank(G)}$ is an abelian subgroup and $N$ is a completion of the subgroup generated by all positive  real root groups ([DGH], [CLL]).

Let $g=k_ga_gn_g\in G(\R)$, written in Iwasawa form. Let $\nu: A^+ \to \mathbb{C}^\times$ be a quasi--character and define
\[
\Phi_{\nu}:G(\R)\to {\mathbb C}^{\times}
\]
to be the function
\[
\Phi_{\nu}(g)=\nu(a_g).
\]
Then $\nu$ is well defined since the Iwasawa decomposition is unique. For convenience, we write $\Gamma=G(\Z)$.

Let $B$ denote the minimal parabolic subgroup of  $G(\R)$.
 Define the Eisenstein series on $G(\R)$ to be the infinite formal sum ([CLL], [CGLLM])
$$E_{\nu}(g)\quad:=\quad \sum_{\gamma\in \Gamma/\Gamma\cap{B}}\quad
\Phi_{\nu}(g\gamma).$$

When $G$ is finite dimensional, we define the {\it constant term} ([CLL], [CGLLM])
$$\int_{{N}/{N}\cap {\Gamma}} E_{\nu}(gu) du$$
where $du$ is a measure on the compact space $N/N\cap\Gamma$. For example, when $G=SL_2$, $\Gamma=SL_2(\Z)$, $N/N\cap\Gamma=S^1$. 

The constant term  corresponds to  the zero--th Fourier
coefficient in the Fourier expansion of the Eisenstein series.

Regarding the complex question of convergence of Eisenstein series, in [CLL], the authors proved convergence of the constant term and hence almost everywhere convergence of the full Eisenstein series on rank $2$ hyperbolic Kac--Moody  groups over $\R$.  The  techniques in [CLL] have been extended to  Kac--Moody groups in general ([CGLLM]).

\subsection{The Weyl group and the constant term}  Let $G$ be a semi simple algebraic group or a Kac--Moody group. Then
 $G$ has Bruhat decomposition $G = \sqcup_{w} G_w$ where
$$
G_w = BwB,
$$
$w\in W$, the Weyl group and $B$ is the `upper triangular', or Borel subgroup. Each Bruhat cell 
$$\Gamma / \Gamma \cap B = \sqcup_{w\in W} (\Gamma\cap G_w) / (\Gamma \cap B)$$ then contributes one term to the constant term. The number of terms in the constant term thus equals the cardinality of the Weyl group $W$.

 When $G$ is a Kac--Moody group, $W$ is infinite.  In [FK], the authors showed that remarkably, for certain choices of the complex parameter $\nu$, there are only finitely many terms in the constant terms for $E_9$, $E_{10}$ and $E_{11}$. This uses  a reduction method of [GMRV] for eliminating terms of the constant term. For $E_9$, this also uses the structure of the affine Weyl group and the Weyl group orbits. This is consistent with the (conjectural) notion that the constant terms of Eisenstein series on Kac--Moody groups should encode a finite number of perturbative string theory
corrections. A detailed analysis of this question is given in [Fl].

Eisenstein series  on non--affine Kac--Moody groups $G(\R)$ are invariant under translations by $G(\Z)$ and hence have a Fourier expansion.  In [CLL], the authors defined and calculated the degenerate Fourier coefficients for Eisenstein series on rank $2$ hyperbolic Kac--Moody  groups over $\R$. In [Fl], the author gave the Fourier integrals needed to obtain the constant term and the higher order Fourier
modes for Eisenstein series in $E_9$, $E_{10}$ and $E_{11}$ and showed how the  `collapse mechanism' of [FK] extends to the higher order Fourier modes. He also gave explicit expressions for the constant terms and Fourier modes of some Kac--Moody Eisenstein series.

\section{Some open questions}
 
 There are certain obstacles to overcome in order to determine the precise role of Kac--Moody Eisenstein series in supergravity. For example, in the study of $E_9(\Z)$ in 2 dimensions,  the supergravity action  cannot be cast in the Einstein frame, that is, the coordinate frame where the coset symmetries are manifest. The objective is to interpret Eisenstein series as coefficients of higher derivative corrections in the string scattering amplitudes. However, without the supergravity action in the Einstein frame, identifying  symmetries in the string scattering amplitudes may be considerably more difficult.  A formulation of the quantum corrections on the level of equations of motion may be required.

The  Kac--Moody Eisenstein series studied in [Fl] have been seen to appear in string scattering amplitudes ([Fl], Sec 2.2.1). Characterizing the underlying physical mechanism at work here will require a more detailed knowledge of string scattering  in $2$ and $1$ dimensions.

 The Eisenstein series should be an eigenfunction of the Casimir operators including the Laplacian on $K(G(\R)) \backslash G(\R)/ G(\Z)$. It would be interesting to find the explicit forms of these eigenvalue equations using the higher order Casimir operators introduced in [K1]. 

 In some cases, certain constraints are required on Eisenstein series defined using representation theory in order that these automorphic forms are eigenfunctions of the Laplacian ([OP1]). It is not yet understood how this might generalize to the Kac--Moody case.

\end{document}